\documentclass[slac_one]{revtex4}
\usepackage{graphicx}
\usepackage{fancyhdr}
\newcommand{\met}{$\not$E$_t$}
\newcommand{\Et}{E$_{\text{T}}$}
\newcommand{\Pt}{P$_{\text{T}}$}

\begin{document}

\title{Production of W and Z bosons accompanied by jets at LHC startup} 

%

\author{Marco De Mattia (for the CMS collaboration)}
\affiliation{INFN and University of Padova, Italy}
\begin{abstract}
We report on potential for measurement of W and Z boson production accompanied by jets at the CMS experiment. Of particular interest are jet multiplicity and \Pt{} distributions. The 10 to 100 pb$^{-1}$ datasets expected in the startup year of operation of LHC are likely to already provide information beyond the reach of the Tevatron collider both in jet multiplicity and \Pt{} range. We are especially interested in understanding the ratios of W+jets to Z+jets distributions by comparing them to next-to-leading order Monte Carlo generators, as these processes present a formidable background for searches of new physics phenomena.
\end{abstract}

\maketitle

\thispagestyle{fancy}


\section{INTRODUCTION} 
Direct production of electroweak gauge bosons (W$^\pm$and Z) in association with jets are important hard processes, that can be used for several studies. First of all, they allow to perform critical tests of perturbative quantum chromodynamic (QCD). The NLO cross sections have been calculated only up to W/Z+2jets and predictions for the higher jet multiplicity final states are accessible through matrix element (+partonshower) computations and in fact can be considered as a prime testing ground for the accuracy of such predictions as well as for measurements of $\alpha_S$. These events can also be used to calibrate the detector, as the jet energy scale systematic uncertainty can be reduced using Z + jet events. The cross section is smaller than for other calibration samples, but the kinematics of the Z boson can be reconstructed very precisely.
W/Z + jets are also an important background for many new particle searches. At the LHC they have high cross sections, with respect e.g. to the Tevatron, and thus constitute the dominant background for a variety of new physics signatures (Higgs, Supersymmetry, ...), besides being a background for SM measurements.
Ultimately, they are a discovery channel for new physics. Any production of new heavy particles with quantum numbers conserved by the strong interaction and EWK couplings is likely to contribute to signatures with one or more EWK gauge bosons; additional jets will always be present at some level from initial-state radiation, and may also be created in cascade decays of new heavy particles or from the decay of associated heavy particles. Precise measurements of the W + jets and Z + jets channels provide a broad search in a number of possible signatures of physics beyond the SM.

\section{THEORY PREDICTIONS}
In this section we will describe theoretical predictions for the production of W/Z+jets at the LHC for the different jet multiplicities \cite{WbosonProductionWithAssociatedJetsAtLargeRapidities, HardInteractionsOfQuarksAndGluons}.

\textbf{\textit{W+1jet}}:
At the LHC the dominant mechanisms for W boson production in association with 1 jet are $qg\rightarrow Wq$ and $q\bar q\rightarrow Wg$, with the first one being the most frequent.
In the $qg\rightarrow Wq$ channel the W is produced abundantly in the forward rapidity region because of the difference in the shape of the p.d.f.’s of the incoming partons.
The same considerations apply also in the case of $q\bar q\rightarrow Wg$ since the $\bar q$ is a sea quark.
In both cases the jet tends to accompany the W in the same rapidity hemisphere or to go in the central region.

\textbf{\textit{W+2jets}}:
The four main subprocesses contributing at LO are: $qg\rightarrow Wqg$, $q\bar q\rightarrow Wgg + Wq\bar q$, $qq\rightarrow Wqq$ and $gg\rightarrow Wq\bar q$. The first one is the dominant process at the LHC.
For the $qg\rightarrow Wqg$ channel, similar considerations as for the W+1jet apply. Due to the p.d.f.s the W boson is produced preferentially in the forward region, with the additional jet favoring even more this configuration. One of the jets tends to follow the W in rapidity. The second jet can be easily separated in rapidity from the W boson.
In $q\bar q\rightarrow Wgg + Wq\bar q$ the kinematical mechanism is the same as in $qg\rightarrow Wqg$ since the antiquark is a sea quark.
The $qq\rightarrow Wqq$ channel is peculiar, since both partons are valence quarks. To make one x large it tends to have the W and one jet slightly forward in rapidity, while to make the other x large it has the second jet well forward and opposite in rapidity.
The $gg\rightarrow Wq\bar q$ channel is perfectly symmetric and the W and the two jets are produced mostly in the central rapidity region.
In most of the cases one of the jets tends to accompany the W in the same rapidity region, while the other jet can be central or in the opposite hemisphere.

\textbf{\textit{W+$>$2jets}}:
There is not direct NLO computation of the cross section in this case. Monte Carlo programs able to simulate this events at LO have been developed and they have proven to be consistent with Tevatron data \cite{TevatronWplusJetsCrossSection}. In particular, the LO predictions and simulation of higher order effects via partonshower generators are already sufficient to describe the data.
It is one of the tasks of the LHC experiments to check whether the same still holds true or if NLO or NNLO approaches are needed.

\textbf{\textit{Z+Njets}}:
At the Tevatron the dominant process for production of Z+jets is $q\bar q\rightarrow$ Z + jets where both partons are valence quarks.
At the LHC the dominant process is the same, but the antiquark comes now from the sea. This means that regions with much smaller x of the p.d.f.s will be explored with Z+Njets events at the LHC.
This channel is not expected to differ much from what described for W+jets and most of the considerations already done in that case apply also for Z+jets.

\section{MEASUREMENT AT LHC STARTUP}
Focus at startup will be on the measurement of the cross sections of W+jets and Z+jets and their ratios with as many accompanying jets as allowed by the statistic. Direct measurements of this cross sections suffer from inherent theoretical and experimental uncertainties associated with the definition and measurement (and hence counting) of jets.
These effects can change the number of jets measured in a given event
 and, together with theoretical uncertainties, mean that the measurement of a specific N-jet exclusive channel at the LHC will be completely dominated by systematic uncertainties.
By considering the ratio $R=(W+Njets)/(Z+Njets)$ most of the uncertainties will cancel (at first order) \cite{RatioAsPreciseTestOfSM}.
Furthermore, most BSM signals can produce deviations in the measured ratio either by decaying into actual Z or W bosons or by having a high missing tranverse energy (\met) signature.
Of course, ratios alone are not very useful to measure the characteristics of the process itself. These require direct measurements on the selected samples.
\subsection{Event Selection}
\begin{figure*}[ht]
 \centering
 \includegraphics[width=0.25\textwidth]{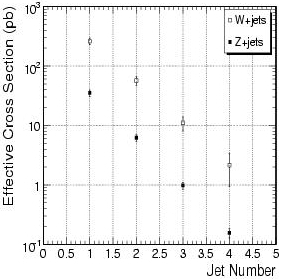}
 \caption{Effective cross section for Z + jets and W +jets events. The effective cross section corresponds to the generated cross section of a physics process with the overall acceptance factor, which includes detector trigger efficiency and offline selection efficiency.}
 \label{fig:effectiveCrossSection}
\end{figure*}

At CMS W/Z+jets events will first be selected by the standard single electron and single muon triggers for the W+jets and the standard double electron and double muon triggers for Z+jets. An offline selection based on the reconstructed leptons and \met{} will be used. Identified leptons with a minimum required transverse momentum are checked for both calorimetric and tracker isolation. This requirement is essential in reducing the huge QCD background. For W identification, a constraint on the transverse mass of the lepton-\met{} system is used. For Z identification a tight cut on the reconstructed mass from the two opposite sign leptons is used.
Jet energy corrections are very important in this kind of analysis. Z+jets events provide the possibility to use reconstructed Z boson momentum to calibrate the jet energy. The same calibration constant set can be used for W+jets, because W+jets and Z+jets have similar jet \Et{} distributions due to the physics correlation between the two processes.
Jet number in the event are counted after jet energy corrections have been applied and a minimum jet transverse energy of about 50 GeV is required. After all these selections the effective cross sections shown in figure \ref{fig:effectiveCrossSection} are found.

\subsection{First Measurements with 10/100 $pb^{-1}$}
With 10pb$^{-1}$, About 2600 (570) W+1jet (2jets) and 350 (60) Z+1jet (2jets) events are expected to be available after selections. This will allow to measure the cross sections and their ratio with 1 and possibly 2 accompanying jets and to test algorithms that use and combine measurements from different detectors (calorimetry, tracker, ...) which are currently being developed. With 100pb$^{-1}$ measurements of the ratio of up to 4 jets and the direct measurement of the cross sections of W+jets and Z+jets are expected. Comparisons with the Monte Carlos will be carried on to assess their predictive capacity especially in the low \Pt{} shapes of the jet energy distributions. These are the most problematic regions and in the long term, once NLO effects are understood, and low \Pt{} shapes well reproduced, systematics can be assigned according to NLO vs NNLO comparisons. Data driven methods will be used wherever possible for the estimation of backgrounds and efficiencies.

After trigger and selection cuts the Z+Njets inclusive samples with up to 4 jets are expected to be at least one order of magnitude above backgrounds.
W+jets events, on the other hand, will still have significant backgrounds. As the number of jets increases, the top background becomes more important; already with 3 jets the background is comparable with the signal and with 4 jets it is even higher than the signal. Top contribution to W+jets events is accentuated at the LHC, since the production rate increases by about a factor of 100 with respect to the Tevatron, while W and Z production increase by just a factor of 5. The reduction of the top background in the W+$>$2 jets channels will be crucial and strategies based on jet characteristics are being developed and tested on Monte Carlo simulations.

\section{CONCLUSIONS}
Among the first measurements that will be carried out at the LHC by the CMS detector there will be the measurement of the cross sections of W+jets and Z+jets and their ratios versus the jet multiplicity. With the first 10pb$^{-1}$ up to W/Z+2 jets are expected to be measured. With the first 100pb$^{-1}$ up to W/Z+4jets measurements will be carried out. Detailed studies of techniques for background subtraction and selection efficiency improvement are ongoing. Monte Carlo generators predictions (both LO and NLO) will be compared to the data with special attention to the low \Pt{} shape of the jet energy distributions.

\end{document}